\documentclass[amsmath,amssymb]{revtex4}

\usepackage{graphicx}
\usepackage{dcolumn}
\usepackage{bm}
\usepackage{amsmath}
\usepackage{amssymb}

\newtheorem{conj}{Conjecture.}

\newcommand{\be}{\begin{equation}}
\newcommand{\ee}{\end{equation}}

\begin{document}

\title{XXX Spin Chain: from Bethe Solution to Open Problems \footnote{ Talk presented at the Solvay workshop BETHE ANSATZ: 75 YEARS LATER
                                                                        October 19-21, 2006, Brussels, Belgium}}

\author{Vladimir E. Korepin}
    \affiliation {C.N. Yang Institute  for Theoretical Physics, State University of New York at Stony Brook, Stony Brook, NY 11794-3840, USA}
       \email{korepin@max2.physics.sunysb.edu}

\author{Ovidiu I. P\^{a}\c{t}u }
   \affiliation{C.N. Yang Institute for Theoretical Physics, State University of New York at Stony Brook, Stony Brook, NY 11794-3800, USA  }
   \altaffiliation{Permanent address: Institute of Space Sciences MG 23, 077125 Bucharest-Magurele, Romania}
       \email{ipatu@grad.physics.sunysb.edu}

\begin{abstract}

Our understanding of Bethe Anstaz has improved a lot over the last 75 years. This was clear from the many excellent lectures
on the conference. But there are still unanswered questions and actually this lecture will concentrate on four open problems. Two of the
problems are related to the correlations functions of the XXX spin chain and the XXZ spin chain one to the entropy of subsystems and
one to the six vertex model with domain wall boundary conditions.

\end{abstract}

\maketitle

\section{Introduction}

Originally introduced in 1931 by Hans Bethe in order to solve the isotropic Heisenberg  spin chain,
the Bethe Ansatz, with its numerous generalizations and refinements has been proven an
invaluable tool in the field of exactly solved models. The versatility
of this method is exemplified by the multitude of physical models and problems in which it proved useful. Some of
these are: $\delta$-function Bose gas, massive Thirring model, Hubbard model, XXX, XXZ and the XYZ spin chains,
six-vertex and eight-vertex models and the list could go on.

Even though our understanding of exactly solved models has improved a lot
there is still a large number of unsolved problems. The purpose of this lecture is
to present some of them.

\section {The Heisenberg  Spin Chain and Number Theory}

Our first open problem is related to the model that was the starting point for the  Bethe Ansatz.
Consider the antiferromagnetic spin 1/2 Heisenberg XXX spin chain with the hamiltonian

\be
\mathcal{H}_{XXX}=\sum_{j=-\infty}^\infty \left( \sigma_j^x\sigma_{j+1}^x+\sigma_j^y\sigma_{j+1}^y+ \sigma_j^z\sigma_{j+1}^z\right)
\ee
where $\sigma_i^x,\sigma_i^y,\sigma_i^z$ are the Pauli matrices and we will denote $\sigma^0$ the $2\times 2$ unit matrix
$$
\sigma^x=\left(\begin{array}{lr}
                0&1\\
                1&0
                \end{array}\right)\ \ \
\sigma^y=\left(\begin{array}{lr}
                0&-i\\
                i&0
                \end{array}\right)\ \ \
\sigma^z=\left(\begin{array}{lr}
                1&0\\
                0&-1
                \end{array}\right)\ \ \
\sigma^0=\left(\begin{array}{lr}
                1&0\\
                0&1
                \end{array}\right)
$$

As we have said this hamiltonian was diagonalized  in 1931 by means of what today we call  coordinate Bethe Ansatz \cite{Bethe}.
The unique antiferromagnetic ground state in the thermodynamic limit was investigated by Hulthen in 1938 \cite{Hulthen} and  the
spectrum of excitations which consists of magnons of spin 1/2 was  correctly described for the first time
by Faddeev and Takhtajan \cite{FT}.

\subsection{Correlation Functions}

The correlation functions of the XXX spin chain are defined as follows: consider $n$ sequential lattice sites and in each site
pick up $\sigma_k^{a_k}\in\{\sigma_k^0,\sigma_k^x,\sigma_k^y,\sigma_k^z\}$. The average of this operator with respect to the ground state
$|GS\rangle$
\be
\langle\prod_{k=1}^n\sigma_k^{a_k}\rangle=\langle GS|\prod_{k=1}^n\sigma_k^{a_k}|GS\rangle
\ee
is called a correlation function. An example is \emph{emptiness formation probability} which is defined as
\be
P(n)=\langle GS|\prod_{j=1}^nP_j|GS\rangle
\ee
where $P_j=(\sigma_j^z+1)/2$ is the projector on the state with the spin up in the $j$-th lattice site. $P(n)$  is the
probability of formation of a ferromagnetic string of length $n$ in the antiferromagnetic ground state.

We will show that there is a strong connection between the correlation functions
of the XXX spin chain and Riemann zeta function at odd arguments, but  first we  need to introduce some
 preliminary notions.

\subsection{Number Theory and Riemann Zeta Function with Odd Arguments}

The roots of all  the polynomials with rational coefficients

$$
r_nx^n+r_{n-1}x^{n-1}+\ldots r_1x+r_0=0,\ \ \ r_i\in\mathbb{Q}
$$
(where we denote by $\mathbb{Q}$ the field of rational numbers ) are called algebraic numbers. The transcendental numbers are not
roots of  any polynomials with rational coefficients (i.e. $\ln 2, \pi, e$). We say that $x,y$ are algebraically dependent if
$$
r_{nm}x^ny^m+\sum_{a=0}^{n-1}\sum_{b=0}^{m-1}r_{ab}x^ay^b=0,\ \ \ \ r_{ij}\in\mathbb{Q}
$$
and algebraically independent if
$$
r_{nm}x^ny^m+\sum_{a=0}^{n-1}\sum_{b=0}^{m-1}r_{ab}x^ay^b\neq0,\ \ \ \ r_{ij}\in\mathbb{Q}
$$
Also  we say that  $t_1,\ldots,t_k$  are algebraically independent transcendental numbers
$$
\sum_{\{a_j\}}r_{a_1\ldots a_k}t_1^{a_1}t_2^{a_2}\ldots t_k^{a_k}\neq 0,\qquad a_j=\mbox{integers},\ \ r_{a_1\ldots a_k}\in\mathbb{Q}
$$
The Riemann zeta function is defined as \cite{Riemann}

\be
\zeta(s)=\sum_{n=1}^\infty\frac{1}{n^s}  \qquad  \Re (s)>1
\ee
and it can also be represented as a product with respect to all the prime numbers $p$ (Euler's product)

\be
\zeta(s)=\prod_{p}\frac{1}{1-p^{-s}}
\ee
For the characterization of the correlation functions  it will also  be useful to use the
alternating zeta series (the value of the polylogarithm at root of unity)
\be
\zeta_a(s)\;=\;\sum_{n>0}{(-1)^{n-1}\over n^s}\;=\; - \mbox{Li}_s(-1)
\label{za}
\ee
where  $\mbox{Li}_s(x)$ is the polylogarithm. The connection between the Riemann zeta function and the alternating zeta series
is
\be
\zeta(s)\;=\;{1\over 1-2^{1-s}}\zeta_a(s)
\label{za1}
\ee
where  $s\neq 1$. Unlike the Riemann zeta function which has a pole at $s=1$ the alternating zeta function
has a limit when $s\rightarrow 1$.
\be
\zeta_a(1)\;=\;\ln 2
\label{zaat1}
\ee

It is known that at even values of the argument the zeta function can be expressed in terms of  powers of $\pi$ but at odd arguments
the situation is more complicated. R. Apery proved first that $\zeta(3)$ is irrational \cite{Ap} and T. Rivoal
showed that the zeta function at odd arguments takes an infinite number of irrational values \cite{Riv} (see also \cite{Zud}).
In fact it was conjectured that the values of the zeta function at odd arguments are all irrational even algebraically
independent transcendent  numbers (Don Zagier \cite{Zagier}, P. Cartier \cite{Cartier}).

\subsection{Quantum Correlations and Number Theory}

In 2001 H.Boos and one of the authors were able to calculate exactly the emptiness formation probability  $P(n)$ for small
strings $(n=1,\ldots,4)$ obtaining the following results

\begin{eqnarray*}
P(1) &=& \frac{1}{2} = 0.5, \nonumber\\
P(2) &=& \frac{1}{3} - \frac{1}{3} \ln 2 \nonumber\\
P(3) &=& \frac{1}{4} - \ln 2 + \frac{3}{8} \zeta(3) = 0.007624158, \nonumber  \\
P(4) &=& \frac{1}{5} - 2 \ln 2 + \frac{173}{60} \zeta(3) - \frac{11}{6} \ln 2 \cdot \zeta(3) - \frac{51}{80} \zeta^2(3)  \nonumber \\
& & \ \ \ \ - \frac{55}{24} \zeta(5) + \frac{85}{24} \ln 2 \cdot \zeta(5) = 0.000206270.\nonumber\\
\end{eqnarray*}

$P(1)$ is obvious from symmetry and $P(2)$ can be obtained from the result of Hulthen \cite{Hulthen}  for  the ground state energy.
 $ P(3)$ can be obtained from Takahashi's result \cite{Tak} for the nearest neighbor correlation (see also \cite{DI}) and $P(4)$ was obtained from
from the integral representation derived in \cite{KIEU} based on the vertex operator approach \cite{MJ}. Subsequent   computations of
$P(5)$ in \cite{BKNS} and $P(6)$ in \cite{BKS} showed that the emptiness formation probability for small strings share the same structure.
This led to the following conjecture

\begin{conj} (Boos, Korepin 2001)
Any correlation function of the XXX spin chain can be represented as a polynomial in $\ln 2$ and values of Riemann
zeta function at odd arguments with rational coefficients \cite{BK},\cite{BKS}.
\end{conj}

The conjecture was sustained also by computations of different correlation functions such as two-point spin-spin correlators
some examples being listed bellow
\begin{eqnarray*}
\left\langle S_j^z S_{j+1}^z \right\rangle
&=& \frac{1}{12} - \frac{1}{3}\zeta_a(1)
 = -0.147715726853315 \nonumber   \\
\left\langle S_j^z S_{j+2}^z \right\rangle
&=&\frac{1}{12} - \frac{4}{3} \zeta_a(1) + \zeta_a(3)
    =  0.060679769956435 \nonumber  \\
\left\langle S_{j}^{z} S_{j+3}^{z} \right\rangle
&=& \frac{1}{12} - 3 \zeta_a(1) + \frac{74}{9} \zeta_a(3) - \frac{56}{9} \zeta_a(1) \zeta_a(3)
    - \frac{8}{3} \zeta_a(3)^2  \nonumber  \\
& & -\frac{50}{9} \zeta_a(5) + \frac{80}{9} \zeta_a(1) \zeta_a(5) \nonumber
    = -0.050248627257235 \\
\left\langle S_{j}^{z} S_{j+4}^{z} \right\rangle
&=& \frac{1}{12} - \frac{16}{3} \zeta_a(1)  + \frac{290}{9} \zeta_a(3) - 72 \zeta_a(1) \zeta_a(3)
   - \frac{1172}{9} \zeta_a(3)^2  - \frac{700}{9} \zeta_a(5)  \nonumber  \\
& &+ \frac{4640}{9} \zeta_a(1) \zeta_a(5) - \frac{220}{9} \zeta_a(3) \zeta_a(5) - \frac{400}{3} \zeta_a(5)^2
     \nonumber  \\
& &+ \frac{455}{9} \zeta_a(7) - \frac{3920}{9} \zeta_a(1) \zeta_a(7)
   + 280 \zeta_a(3) \zeta_a(7)  \nonumber \\
&=&  0.034652776982728 \nonumber
\end{eqnarray*}

Again the nearest neighbor correlator was obtained from Hulthen result \cite{Hulthen} and the second-neighbor
correlator was obtained by Takahashi in 1977 \cite{Tak} using the strong coupling expansion of the ground state energy
of the half-filled Hubbard chain. The next nearest correlators were calculated in \cite{SSNT},\cite{BST}.

The conjecture was finally proved in 2006   by H. Boos, M. Jimbo, T. Miwa, F. Smirnov and Y. Takeyama \cite{BJMST}.
However we still not have explicit expressions for  the rational coefficients that enter in the expression for the
correlation functions as polynomials in ln 2 and Riemann zeta function at odd arguments.
Such a description will provide a tractable formula for the correlations of the XXX spin chain.

{\bf  Open problem:  Efficient description of the rational coefficients which appear in the expression for the correlation functions of the XXX
spin chain as a polynomial in alternating zeta series at odd arguments.}


\section{Entropy of Subsystems}

We are interested in the following physical situation. Consider a one-dimensional system of interacting spins (we can also consider
particles interacting) in the ground-state
denoted by $|GS\rangle.$ We will treat the whole ground state as a binary system $A$ and $B$  where $A$ is a block of neighboring
spins and $B$ is the rest of the  spins in the ground state. The density matrix of the entire system is
\be
\rho_{A\&B}=|GS \rangle \langle GS|
\ee
and the density matrix of the subsystem $A$  obtained by tracing away the $B$ degrees of freedom is
\be
\rho_A=Tr_B(\rho_{A\&B})
\ee
In a seminal paper \cite{BBPS} Bennet, Bernstein, Popescu and Schumacher discovered that the von Neumann entropy of a  subsystem
$A$
\be\label{es}
S(\rho_A)=-Tr(\rho_A\ln\rho_A)
\ee
is a measure of entanglement. Entanglement is the fundamental resource used in quantum computation and quantum information. A better
understanding of entanglement will provide further insight in the theory of quantum phase transitions but also in the physics of
strongly correlated quantum systems where  deeply entangled ground-states play a  major role in the understanding of these quantum
collective phenomena. Consequently a large amount of effort was invested in studying the entropy of subsystems in a large
class of quantum systems.

\subsection{General results}

If we consider the doubling scaling limit in which the size of the block of spins  is much larger than one but much smaller than the length of the entire
chain we can present some general results about the behavior of the entropy of subsystems.

In the case of 1D critical models (gap-less) the entropy of the subsystems scales logarithmically  with the size of the block. More precisely
for a block of $n$ spins we have

\be\label{se}
S(n)=\frac{c}{3}\ln n\ \ \ \ \ n\rightarrow\infty
\ee

where $c$ is the central charge of the associated conformal field theory that describes the critical model. This formula was first derived
for the geometrical entropy (the analogous of  (\ref{es}) for conformal field theory) by Holzhey, Larsen and Wilczek in
\cite{HLW} (see also \cite{VLRK},\cite{K},\cite{CC}). Some examples are: Hubbard model , XX0 (or isotropic XY) spin chain, higher spin
generalization of the isotropic XXX antiferromagnetic spin chain, Bose gas with $\delta$ interaction etc.

In the case of non-critical models (gap-full) it was conjectured in \cite{VLRK} (based on numerical evidence for some spin chains)
that the entropy of subsystems will increase  with the size of the subsytem  until it will reach a limiting value $S(\infty)$.
This was proved for the XY spin chain in \cite{IJK},\cite{IJK2} where the limiting value was analytical computed and also checked for
the spin chain introduced by Affleck, Kennedy, Lieb, Tasaki
(\cite{AKLT} AKLT model) in \cite{FKR} where it was showed that $S(\infty)=2$. This explains why the Density Matrix Renormalization Group \cite{W}(DMRG)
technique works so well for the AKLT model and fails to reproduce quantum critical behavior. It was noted in \cite{VLRK} that in order for the DMRG
technique to work we need to have for the $S(\rho_A)$ a limiting value as the size of the subsystem increases (this is equivalent with a bounded rank for
$\rho_A$) and as we have shown this is not the case in critical models.

\subsection{An example: the XY spin chain}

We can make all these general considerations more exact by presenting the results for the XY spin chain in magnetic field with the hamiltonian

$$
\mathcal{H}_{XY}=-\sum_{j=-\infty}^\infty\left((1+\gamma) \sigma_j^x\sigma_{j+1}^x+(1-\gamma)\sigma_j^y\sigma_{j+1}^y+h\sigma_j^z\right)
$$
where $0<\gamma< 1$ is the anisotropy parameter an $h$ is the magnetic field.
The model was solved   by E.H. Lieb, T. Schulz and D. Mattis  in zero magnetic  field case \cite{LSM} and by E. Barouch and B.M. McCoy in the
presence of a constant magnetic field \cite{BM}. The ground state is unique and in general there is  a  gap in the spectrum.

The density matrix of a  block of $n$ neighboring spins in the ground state can  be expressed as

$$
\rho (n)=\frac{1}{2^n}\sum_{\begin{subarray}{c}\{a_j\} \\ j=1..n \end{subarray}}\left( \prod_{j=1}^n\sigma_j^{a_i} \right)<GS|\prod_{k=1}^n\sigma_k^{a_k}|GS>
$$
and the limiting value of the entropy  in the double scaling limit depends on the isotropy and magnetic field.
We can distinguish three cases:
\begin{itemize}
\item Case Ia: moderate magnetic field $2\sqrt{1-\gamma^2}<h<2$
\item Case Ib: weak magnetic field including zero magnetic field $0\leq h<2\sqrt{1-\gamma^2}$
\item Case II: strong magnetic field $h>2$
\end{itemize}
The result for  these regions obtained in \cite{IJK} (we denote by  $S(\infty)$ the limiting value of  $S(\rho(n))$ when $n\rightarrow\infty$) is
$$
S(\infty)=\frac{\pi}{2}\int_0^\infty\ln\left(\frac{\theta_3(ix+\frac{\sigma\tau}{2})\theta_3(ix-\frac{\sigma\tau}{2})}{\theta_3^2(\frac{\sigma\tau}{2})}
\right)\frac{dx}{\sinh^2(\pi x)}
$$
where the modulus $k$ of the theta function is different in the three regions; $\tau=I(k')/I(k)$ where $I(k)$ is the complete elliptic integral
of modulus $k$, $k'=\sqrt{1-k^2}$ is the complementary modulus and $\sigma=1$ in Case I and $\sigma=0$ in Case II.
Using the approach of \cite{CC}  I. Peschel independently calculated and simplified the results in region (Ia) and (II) obtaining
$$
S(\infty)=\frac{1}{6}\left[\ln\left(\frac{k^2}{16k'}\right)+\left(1-\frac{k^2}{2}\right)\frac{4I(k)I(k')}{\pi}\right]+\ln 2
\ \ \ \  \mbox{ with }\ \ \ \  k=\frac{\sqrt{(h/2)^2+\gamma^2-1}}{\gamma}\ \ \ \ \mbox{ (Ia) }
$$
$$
S(\infty)=\frac{1}{12}\left[\ln\frac{16}{k^2k'^2}+(k^2-k'^2)\frac{4I(k)I(k')}{\pi}\right]\ \ \ \ \mbox{ with }\ \ \ \
k=\frac{\gamma}{\sqrt{(h/2)^2+\gamma^2-1}}\ \ \ \ \mbox{ (II) }
$$
and the  simplified result for the region (Ib) was finally obtained in \cite{IJK}
$$
S(\infty)=\frac{1}{6}\left[\ln\left(\frac{k^2}{16k'}\right)+\left(1-\frac{k^2}{2}\right)\frac{4I(k)I(k')}{\pi}\right]+\ln 2
\ \ \ \  \mbox{ with }\ \ \ \  k=\frac{\sqrt{1-\gamma^2-(h/2)^2}}{\sqrt{1-(h/2)^2}}\ \ \ \ \mbox{ (Ib) }
$$

In the isotropic case $\gamma=0$ the ground-state is again unique but now the model is critical for $h<2$. We expect that the entropy will scale
logarithmically and indeed it was showed in \cite{JK} that

$$
S (\rho (n))=\frac{1}{3}\ln (n\sqrt{4-h^2}) -\int_{0}^\infty dt\left\{\frac{e^{-3t}}{3t}+\frac{1}{t\sinh^2(t/2)}-\frac{\cosh{t/2}}{2\sinh^3(t/2)}\right\}
$$
We see that the result of applying  the magnetic field  is very simple  effectively reducing the size of the subsystem. If the magnetic
field is larger than the critical value $2$ then the ground-state is ferromagnetic and the entropy is zero.

\subsection{The  XXZ spin chain}

It will be highly desirable if we would be able to obtain the same amount of information about the
XXZ spin chain with the hamiltonian

$$
\mathcal{H}_{XXZ}=-\sum_{j=-\infty}^{\infty}\left( \sigma_j^x\sigma_{j+1}^x+\sigma_j^y\sigma_{j+1}^y+\Delta \sigma_i^z\sigma_{j+1}^z\right)
$$

If $\Delta>1$ the ground-state is ferromagnetic so $S(\rho(n))=0$ and in the critical region ($-1\leq\Delta<1$) the entropy will scale logarithmically (see \ref{se}).
In the gapped antiferromagnetic case ($\Delta<-1$) we expect that $S(\rho(n))$ will tend  to a limiting value $S(\infty)$ but at this
moment the analytic evaluation of this constant is missing.

{\bf Open problem: For the XXZ spin chain in the antiferromagnetic region ($\Delta<-1$) calculate the limiting value of the entropy
of a subsystem when the number of spins in the block is large.}

%

\section{ Asymptotics of Time and Temperature Dependent Correlation Functions }

The third open problem that are we going to present is related to the asymptotic behavior of correlations functions when the space and time
separation is large. As in the previous section we are going to present two examples for which we have satisfactory  results
and which we believe are going to facilitate the understanding of the problem at hand.

Consider the isotropic XY model \cite{LSM}  in transverse magnetic field with the hamiltonian
$$
\mathcal{H}_{XY}=-\sum_{j=-\infty}^{\infty}\left( \sigma_j^x\sigma_{j+1}^x+\sigma_j^y\sigma_{j+1}^y+h\sigma_j^z\right)
$$
where $\sigma$ are Pauli matrices and $h$ is the magnetic field. We are interested in the asymptotic behavior of the time
and temperature correlation function
\be\label{cf}
g(n,t)=\frac{Tr\ \left\{ \left( e^{-\mathcal{H}_{XY}/T}\right) \sigma_{n_2}^+(t_2) \sigma_{n_1}^-(t_1)\right\}}{Tr\ \left(  e^{-\mathcal{H}_{XY}/T}\right)}
\ee
when $n=n_2-n_1$ and $ t=t_2-t_1$ are large and $ h\in[0,2).$ In \cite{IIKS} it was showed that $g(n,t)$ decays exponentially but the rate of decay depends
on the direction $\phi$ defined as $\cot \phi=n/4t$ when $n,t\rightarrow\infty.$ The asymptotics in the space-like and time-like regions are:
\begin{itemize}
\item Space-like directions $0\leq\phi<\pi/4$
\be
g(n,t)\rightarrow C\exp\left\{\frac{n}{2\pi}\int_{-\pi}^{\pi}dp\ \ln\left|\tanh\left[\frac{h-2\cos p}{T}\right]\right|\right\}
\ee
\item Time-like directions  $  \pi/4<\phi\leq\pi/2 $
\be
  g(n,t)\rightarrow Ct^{(2\nu_+^2+2\nu_-^2)}\exp\left\{\frac{1}{2\pi}\int_{-\pi}^{\pi}dp\ |n-4t\sin p|\right.
 \left.\ln\left|\tanh\left[\frac{h-2\cos p}{T}\right]\right|\right\}
\ee
with
$$
\nu_+=\frac{1}{2\pi}\left|\tanh\left(\frac{h-2\cos p_0}{T}\right)\right|\ \ \ \ \ \ \  \nu_-=\frac{1}{2\pi}\left|\tanh\left(\frac{h+2\cos p_0}{T}\right)\right|
\ \ \ \ \ \ \ \frac{n}{4t}=\sin p_0
$$
\end{itemize}

At zero temperature the asymptotics of the correlation functions were evaluated in \cite{MPS} and \cite{VT} and for $\phi=\pi/2$ the leading factor
was computed in \cite{DZ}.

Similar formulae were also obtained for the $\delta$-function Bose  gas ($N$ bosons interacting via a repulsive $\delta$-function
potential of strength $c$) which is characterized by the hamiltonian

\be
\mathcal{H}_N=-\sum_{j=1}^N\frac{\partial^2}{\partial x_j^2}+2c\sum_{N\geq j>k\geq 1}\delta(x_j-x_k)
\ee
In this case the asymptotic formula ($x,t\rightarrow\infty$) obtained in \cite{KS} using the determinant approach to quantum correlation
functions \cite{bki} is
\be
\langle\psi(0,0)\psi^\dagger(x,t)\rangle_T \longrightarrow\exp\left\{\frac{1}{2\pi}\int_{-\infty}^\infty\frac{d\lambda}{2\pi\rho_t(\lambda)}
               |x-v(\lambda)t|\ln\left|\tanh\left(\frac{\epsilon(\lambda)}{2T}\right)\right|\right\}
\ee
where $\epsilon(\lambda), \rho_t(\lambda) $ and $v(\lambda)$ are the finite temperature dressed energy, pseudoparticle density and Fermi velocity.

\subsection{The  XXZ Spin Chain}

In the last years remarkable progress was made in obtaining multiple integral representation for the
correlation functions of the XXZ spin chain. We remind the hamiltonian

$$
\mathcal{H}_{XXZ}=-\sum_{j=-\infty}^{\infty}\left( \sigma_j^x\sigma_{j+1}^x+\sigma_j^y\sigma_{j+1}^y+\Delta \sigma_i^z\sigma_{j+1}^z+h\sigma_j^z\right)
$$
where $\Delta$ is the anisotropy and $h$ the magnetic field. Using the q-vertex operator approach  and corner transfer matrices,
Jimbo, Miki, Miwa and Nakayashiki  obtained in 1992 \cite{JMMN} multiple integral representation for the correlation functions
in the massive regime ($\Delta<-1$), and a conjecture for the critical regime ($|\Delta|< 1$) was proposed in \cite{JM}.
The next important steps were made by Kitanine, Maillet, Terras and later Slavnov
when they proved the previous results and their extension in the case of a magnetic field in both regimes in 1999 \cite{KMT1,KMT2}.
Their method used the algebraic Bethe Ansatz \cite{bki}. 
 Later they were able to obtain representations for the two-point correlations in terms of a  multiple integral  \cite{KMST1} which
they called \emph{master equation}. The generalization of the method for the time-dependent correlations was made in \cite{KMST2} (see \cite{KMST3} for a review of the entire series of papers). The final step was the extension of the multiple integral representation
for finite temperatures which was done by G\"ohmann, Kl\"umper, Seel and Hasenclever in \cite{GKS,GHS}. It should also be mentioned that
in the study of the correlation functions in integrable models an important role is played by the quantum Knizhnik-Zamolodchikov equation
\cite{KZ} (see also \cite{FR},\cite{S}) and for supersymmetric fermion models determinant representation of the correlations functions
were reviewed in \cite{ZYZ}.

Despite these considerable efforts the asymptotics  of time and temperature dependent correlation functions, such as
$g(n,t)$ (see \ref{cf}) for the XXZ spin chain,  are still out of our reach. We expect that they will decay exponentially like in the previous
examples of the isotropic XY spin
chain or the $\delta$-function Bose gas but at this moment we do not have such a result.

{\bf Open problem: Derive an  explicit analytic formula describing the asymptotic exponential decay of time and temperature correlation
functions in  the critical region ($-1\leq\Delta< 1$) of the XXZ spin chain.}

\section{Six Vertex Model with Domain Wall Boundary Conditions}

The six vertex model with domain wall boundary conditions (DWBC) was introduced in \cite{K1} where recursion relations for the
partition function were also derived. By domain wall boundary conditions we understand the situation in which the arrows
on the upper and lower boundaries point in the square lattice  and the ones on the left and right boundaries point out.

Using the recursion relations Izergin was able to derive a determinantal formula for the partition function. In the case
of the homogeneous system if we parametrize the weights as

\be
a=\sin(\gamma-t)\ \ \ \ b=sin(\gamma+t)\ \ \ \ c=sin(2\gamma),\ \ \ \ \ \ |t|<\gamma
\ee
then the result obtained in \cite{I} is
\be\label{IK}
Z_N=\frac{[\sin(\gamma+t)\sin(\gamma-t)]^{N^2}}{(\prod_{n=0}^{N-1}n!)^2}\tau_N
\ee
where $\tau_N$ is the H\"{a}nkel determinant
\be
\tau_N=\det\left(\frac{d^{i+k-2}\phi}{dt^{i+k-2}}\right)_{1\leq i,k\leq N}
\ee
and
\be
\phi(t)=\frac{\sin(2\gamma)}{\sin(\gamma+t)\sin(\gamma-t)}
\ee

\subsection {Connection to Algebraic Combinatorics}

One important feature of the six vertex model with DWBC on a $N$ by  $N$ lattice is the connection with the alternating sign matrices of the
same dimension. An alternating sign matrix (ASM) is a matrix in which all the entries are $-1,0,1$, the sequences of $1$'s and $-1$'s
are alternating and the sum of all the elements  on each row and column is 1. An example of dimension 4 is

\be
\left(\begin{array}{llll}\ 0 &\ 1&\ 0&\ 0\\
                       \ 0 &\ 0&\ 1&\ 0\\
                       \ 1 &-1&\ 0&\ 1\\
                       \ 0 &\  1&\  0& \ 0
       \end{array}\right)
\ee
The study of ASM is the domain of algebraic combinatorics.  The number of ASM of dimension $N$ was conjectured in \cite{MRR1,MRR2}
to be
\be
A(N)=\prod_{n=0}^{N-1}\frac{(3n+1)!n!}{(2n)!(2n+1)!}
\ee
and it was proved for the first time by Zeilberger \cite{Zeil}. Using the fact that there is a one-to-one connection between the allowed configurations
of the six vertex with DWBC and ASM and the determinantal formula (\ref{IK}) Kuperberg was able to give an alternative proof \cite{Kup}
 to the ASM conjecture (see also \cite{B}). An unified and simplified treatment of ASM  enumerations and the relation with the classical
 orthogonal polynomials was found by Colomo and Pronko in \cite{CP}.

\subsection {The Thermodynamic Limit }

From the physicist's point of view a more important feature of the
six vertex model with DWBC is the dependence of the bulk free energy
on the boundary conditions. The fact that particularly  boundary
conditions affect the bulk free energy of the six vertex model was
discovered in \cite{E}. The study of the thermodynamic limit of the
bulk free energy for the six vertex model with DWBC was initiated in
\cite{KZJ}. Using the determinant representation of the partition
function a Toda differential equation was asymptotically solved in
order to extract the bulk free energy in the disordered and the
ferroelectric phases. It was noticed that for the disordered region
the result is different  from the result derived for periodic
boundary conditions but the phase transitions take place in the same
place. In \cite{ZJ} P. Zinn-Justin was able to rewrite the
determinantal formula in terms of a partition function of a random
matrix model. This allowed him to obtain also the bulk free energy
in the antiferroelectric phase and to conjecture that in the
disordered region the partition function behaves like

\be\label{kappa}
Z_N\sim C N^{\kappa}e^{N^2f}\ \ \mbox{ as }\ \
N\rightarrow\infty
\ee
with $C$ a constant and the bulk free energy is
\be
f=\ln\left(\frac{\pi[\cos(2t)-\cos(2\gamma)]}{4\gamma\cos\left(\frac{\pi
t}{2\gamma}\right)}\right) \ee

Using the results from \cite{ZJ}, Bleher and Fokin obtained the large $N$ asymptotics of the free energy in the disordered region \cite{BF}.
They have used completely integrable integral operators, the Riemann-Hilbert approach and the Deift-Zhou nonlinear steepest descent method.
 This allowed Bleher and Fokin  to prove Zinn-Justin's conjecture obtaining for the $\kappa$ exponent (see \ref{kappa}) the value

\be
\kappa=\frac{1}{12}-\frac{2\gamma^2}{3\pi(\pi-2\gamma)}
\ee

\subsection {Correlation Functions}

As in all exactly solvable models the study of correlation functions is of extreme importance. In the case of the six vertex model
with DWBC an added difficulty is constituted by the lack of translation invariance. However in the vicinity of the boundary
some results are known. Determinant representations for the 1-point boundary correlation function were derived in \cite{BPZ} using
the algebraic Bethe Ansatz. This result was rederived using a direct combinatorial approach and extended to 2-point boundary correlation functions
in \cite{FP}. In principle this technique can be used to express more general boundary correlation functions as sums over determinants.
For a different calculation  of the 2-point boundary correlation function and the relation with the doubly refined $x$-enumerations of ASM see \cite{CP1}.

{\bf Open problem: Calculate the bulk and boundary correlation functions of the six vertex model with domain wall boundary
conditions. }

\subsection {Further Generalizations}

The main feature of formula  (\ref{IK}) is the fact that the partition function of the six vertex model with DWBC
is expressed as a determinant of the size of the system. The next obvious step would be the generalization of DWBC in the case of
other models such that the partition function share the same feature. 
 In the case of the six vertex model  the generalization for spin-$k$/2 was done in \cite{CFK} using a fusion procedure, for the
level-1 \emph{so(n)} affine vertex models the generalization was made in \cite{DF} and recently DWBC were obtained for supersymmetric models \cite{ZZ2}.

{\bf Open problem: Obtain generalizations of domain wall boundary conditions for other models such that the partition function
     is expressed as a determinant of a dimension  proportional with the size of the system. }

\section{Conclusions}

Of course the open problems that we have presented do not exhaust the list of unanswered questions in the field of exactly solvable models.
We should mention for example the solving of the time-dependent Schr\"{o}dinger equation for spin chains which plays  an important role
in quantum information theory \cite{CE}. In the light of the talks given at the conference we can definitely say  that
Bethe Ansatz is alive and well after 75 years of developments and will carry on into the 21st century.

\end{document}